\documentclass[twocolumn, aps, pra, 10pt]{revtex4-2}

\usepackage{graphicx}

\usepackage{amsmath, amssymb, amsfonts}

\usepackage{hyperref}
\usepackage{xcolor}

\begin{document}

    \title{Generation of bipartite mechanical cat state by performing projective Bell state measurement on a pair of superconducting qubit.}
    \author{Roson Nongthombam}
        \email{n.roson@iitg.ac.in}
        \affiliation{Department of Physics, Indian Institute of Technology Guwahati, Guwahati-781039, India}
    \author{Urmimala Dewan}
        \email{d.urmimala@iitg.ac.in}
        \affiliation{Department of Physics, Indian Institute of Technology Guwahati, Guwahati-781039, India}
    \author{Amarendra K. Sarma}
        \email{aksarma@iitg.ac.in}
        \affiliation{Department of Physics, Indian Institute of Technology Guwahati, Guwahati-781039, India}


    \begin{abstract}
      Quantum state preparation and measurement of photonic and phononic Schrödinger cat states have gathered significant interest due to their implications for alternative encoding schemes in quantum computation. These scheme employ coherent state superpositions, leveraging the expanded Hilbert space provided by cavity or mechanical resonators in contrast to two-level systems. Moreover, such cat states also serve as a platform for testing fundamental quantum phenomena in macroscopic systems.
    In this study, we generate four bipartite phononic cat states using an entanglement swapping scheme achieved through projective Bell state measurements on two superconducting qubits. Employing two superconducting qubits allows for the creation of bipartite phononic cat states remotely, where the two phononic resonators are separated by a far distance. Subsequently, we conduct a Bell inequality test on the bipartite cat state using the CHSH formulation. Given that the entangled cat states are generated through entanglement swapping, our approach holds promising applications for the advancement of complex quantum network processors based on continuous variable systems.
    \end{abstract}

    \maketitle


    \section{Introduction}
        \label{sec:intro}
        Observing the quantum behaviour of macroscopic mechanical structures continues to be a challenging endeavour and a significant step towards realizing quantum-acoustic state preparation and measurement. The intriguing aspect of these structures, being massive yet displaying quantum behaviour, makes them important platforms for implementing various quantum technological applications and for exploring fundamental physics questions. Recent experiments have unequivocally demonstrated the quantum properties of solid-state mechanical objects \cite{10.1038nature08967, Nature.571.537–540.(2019), Nature.563.666–670.(2018), SCIENCE7.372.6542.622-625, Nature.556.478–482.(2018), SCIENCE.370.6518.840-843}. These include interfacing mechanical objects with the strong quantum nonlinearity of superconducting qubits, leading to the field of circuit quantum acoustodynamics (cQAD)  \cite{ Nature.494.211–215.(2013), NaturePhys.11.635–639.(2015), Science364.368-371(2019),  Nature.459.960–964.(2009), PRXQuantum.4.040342, Nature.604.463–467.(2022),Nat.Phys.18.94–799.(2022)} which is analogous to the well-developed field of circuit quantum electrodynamics (cQED) \cite{RevModPhys.93.025005, Nature.431.162–167.(2004), PhysRevA.75.032329}. The integration of superconducting qubits serves as a quantum-acoustic state preparation and measurement element for the mechanical system \cite{Nature.604.463–467.(2022), Nat.Phys.18.794–799.(2022)}. These state preparations and measurements are some of the basic building blocks for constructing acoustic quantum memories and processors.
        
        A very significant quantum-acoustic state that can be prepared from the superconducting qubit-mechanical integrated system is the phononic Schrödinger cat state, defined as quantum superpositions of quasi-classical coherent states.
        Such phononic cat state preparation has been studied in \cite{npjQuantumInf.8.74(2022), Science380.274-278(2023)}. Similar photonic cat states have also been studied in other various quantum systems such as circuit quantum electrodynamics \cite{Science342.607-610(2013),Science352.1087-1091(2016)}, vibrational states of trapped ions \cite{RevModPhys.85.1103}, propagating photon modes \cite{NaturePhoton.13.110–115(2019), Science312.83-86(2006)}. Preparing cat states has attracted wide research attention owing to its applications in implementing quantum metrology \cite{PhysRevA.100.032318} and quantum information processing protocols based on continuous-variable cat states \cite{PhysRevX.9.041053}, as well as testing fundamental quantum phenomena in macroscopic systems \cite{NatCommun.6.8970.(2015)}. One of the most fundamental tests of quantum phenomena in a quantum system is the Bell inequality test \cite{Nature.526.682–686(2015), PhysRevLett.115.250401, PhysRevLett.115.250402, Nature.617.265–270.(2023)}. Bell inequality tests, performed on pairs of spatially separated entangled quantum systems, demonstrate that quantum systems do not adhere to the principle of local causality. In this work, we use the Clauser–Horne–Shimony–Holt (CHSH) \cite{PhysRevLett.23.880} formulation of the Bell test to conduct the Bell inequality test on a bipartite entangled phononic cat state generated through projective Bell state measurement on a pair of superconducting qubits \cite{PhysRevLett.123.060502}. The creation of bipartite photonic cat states using a single qubit is demonstrated in \cite{Science352.1087-1091(2016), Sci.Adv.8.eabn1778}. In our scheme, 
        four bipartite phononic cat states, or four phononic Bell states, are generated on two phononic 
        crystal mechanical resonators, interacting piezoelectrically with a pair of superconducting qubits via capacitive coupling. The two qubits are connected through a microwave cavity resonator, and through virtual excitation of the cavity photon, they become entangled, creating qubit Bell state. Upon projective measurement of the Bell state of the qubits, each of the four bipartite phononic cat states can be distinguished. This measurement effectively swaps the entanglement from qubit-mechanical to mechanical-mechanical pairs. Such entanglement swapping schemes are pivotal in quantum repeaters, essential for realizing long-distance quantum communication and complex quantum networks \cite{JOSAB.27.00A137, SciRep.13.21998(2023),PhysRevLett.128.150502}. Since the two qubits are connected via a microwave cavity bus, the phononic crystal resonator can be placed at far ends. This setup allows for the creation of bipartite phononic cat states remotely, enabling long-distance quantum state preparation. 
        By harnessing this capability and employing entanglement swapping schemes, the bipartite phononic cat states generated in this study hold promise for practical applications in implementing quantum network processors relying on continuous variable resonators. 

        We begin in Section II with a brief introduction of the hybrid electromechanical system under study and then discuss the generation of resonator-qubit Bell cat state.
        In Section III, we examine the projective Bell state measurement on the two qubits, which leads to entanglement swapping from qubit-qubit to resonator-resonator. We show how this Bell state measurement is realized and how four bipartite phononic Bell cat states are created as a result of the projective measurement. Moving on to Section IV, we conduct the Bell inequality test on the bipartite cat state using the CHSH formulation.


    \section{Qubit-mechanical resonator entanglement.}
        \label{sec:system}

\begin{figure}[ht]
            \centering
            \includegraphics[width=0.47\textwidth]{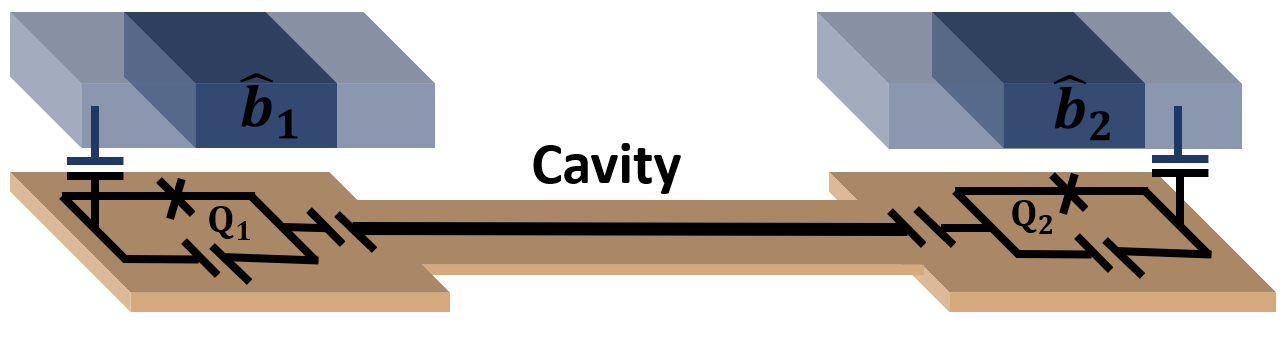}
            \caption{(Color online) Schematic of the hybrid qubit-resonator system. Two qubits denoted by $Q_1$ and $Q_2$ are capacitively coupled to the phononic crystal resonators $\hat{b}_1$ and $\hat{b}_2$, respectively \cite{Nature.604.463–467(2022), PRXQuantum.4.040342}. The qubits interact piezoelectrically with the resonators. The two qubits are coupled to each other via a microwave cavity resonator. This coupling enables the interaction between the two qubits by exchanging virtual excitations of the cavity photons.}
            \label{fig:1}
        \end{figure}

        We consider two hybrid electromechanical systems, each comprising a mechanical resonator dispersively coupled to a superconducting qubit as shown in Fig. \ref{fig:1}. Assuming that the two qubit-mechanical pairs are uncoupled (appendix \ref{app:ham_dispersive}), the Hamiltonian of the two hybrid systems reads as
         
        \begin{subequations}
            \begin{eqnarray}
                \label{eqn:ham_qm}
                \hat{H}_{bq1} &=& \frac{\hbar}{2} \Omega_1 |e_1\rangle\langle e_1| + \frac{\hbar \omega_1}{2}\hat{b}_1^{\dagger}\hat{b}_1 + \hbar\lambda_1\,\hat{b}_1^{\dagger}\hat{b}_1\,|e_1\rangle\langle e_1| ,  \\ 
                \label{eqn:ham_om}
                \hat{H}_{bq2} &=& \frac{\hbar}{2} \Omega_2 |e_2\rangle\langle e_2|  + \frac{\hbar\omega_2}{2}\hat{b}_2^{\dagger}\hat{b}_2 + \hbar\lambda_2\,\hat{b}_2^{\dagger}\hat{b}_2\,|e_2\rangle\langle e_2| .
            \end{eqnarray}
        \end{subequations}
        Here, $\Omega_1$ ($\Omega_2)$ and $\omega_1$ $(\omega_2)$ are the qubit and mechanical frequency of the first (second) hybrid system. $\hat{b}_1$ and $\hat{b}_2$ are the operators of the two mechanical resonators. In the dispersive coupling, the detuning $\delta_1 = \omega_1-\Omega_1$ ($\delta_2 = \omega_2-\Omega_2$) is much larger than the resonant coupling strength between the qubit and the mechanical resonator. We prepare two entangled Bell-cat states by evolving the Hamiltonian $\hat{H}_{bq1}$ and $\hat{H}_{bq2}$. Initiating the qubits in the superposition state and the mechanical resonators in the coherent state, the state of the first hybrid system in the interaction frame after some time $t$ becomes
        \begin{equation}
        \label{eqn:2}
        |\psi\rangle_1 = (|\beta_1 e^{i \lambda_1t}\rangle|e_1\rangle + |\beta_1 \rangle|g_1\rangle)/\sqrt{2},
        \end{equation}
        where $|\beta_1\rangle$ is the coherent state of the first mechanical resonator, while $|e_1\rangle$ and $|g_1\rangle$ refer to the excited and the ground states of the first qubit, respectively.
        We get a similar state $|\psi\rangle_2$ for the second hybrid system.
        At time interval $t=(2n-1)\pi/\lambda_1$, where $n=1,2,3...$, we get the Bell-cat state of the qubit-mechanical system.
        
        The fidelity and entanglement of the bipartite state $|\psi\rangle_1$ in the presence of a noisy environment is shown in Fig. \ref{eqn:2}. The noisy environment is included by solving the Lindblad master equation.
        \begin{eqnarray}
        \label{eqn:3}
        \dot{\hat{\rho}}_1 &=& -\frac{i}{\hbar} [ \hat{H}_{int},\hat{\rho}_1 ] + \gamma_1(n_1+1) \mathcal{L} [ \hat{b}_1 ] + \gamma_1n_1\mathcal{L} [ \hat{b}_1^\dagger ]\nonumber\\
        && + \Gamma_1 \mathcal{L} \left[ \hat{\sigma}_- \right] + \Gamma_1 \mathcal{L} \left[ \hat{\sigma}_+ \right]+\Gamma_1 \mathcal{L} \left[ \hat{\sigma}_z \right],
        \end{eqnarray}
        where $\mathcal{L} [ \hat{o} ] = ( 2 \hat{o} \hat{\rho} \hat{o}^{\dagger} - \hat{o}^{\dagger} \hat{o} \hat{\rho} - \hat{\rho} \hat{o}^{\dagger} \hat{o} ) / 2$ with $\hat{o} \in \{ \hat{b}_1, \hat{\sigma}_-,\hat{\sigma}_z \}$.
        $\gamma_1$ and $\Gamma_1$ are the decay rates of the mechanical resonator and the qubit, respectively.
        The entanglement of the qubit-mechanical bipartite system is computed using the relation $E_N(\hat{\rho}_1)=log_2||\hat{\rho}^{T_A}_1||$, where $\hat{\rho}^{T_A}_1$ is the trace norm of the partial transpose of the bipartite mixed state $\hat{\rho}_1$ \cite{Phys.Rev.Lett.95.090503,PhysRevA.65.032314}. As shown in the figure, the fidelity (F) reaches near one at the interval $t=0.39\mu s$, $1.18\mu s$, and so on for coupling constant $\lambda_1= 8$MHz. Therefore, the qubit-mechanical bipartite system evolves into a Bell-cat state at the interval of $\pi$. 
        
        \begin{figure}[ht]
            \centering
            \includegraphics[width=0.3\textwidth]{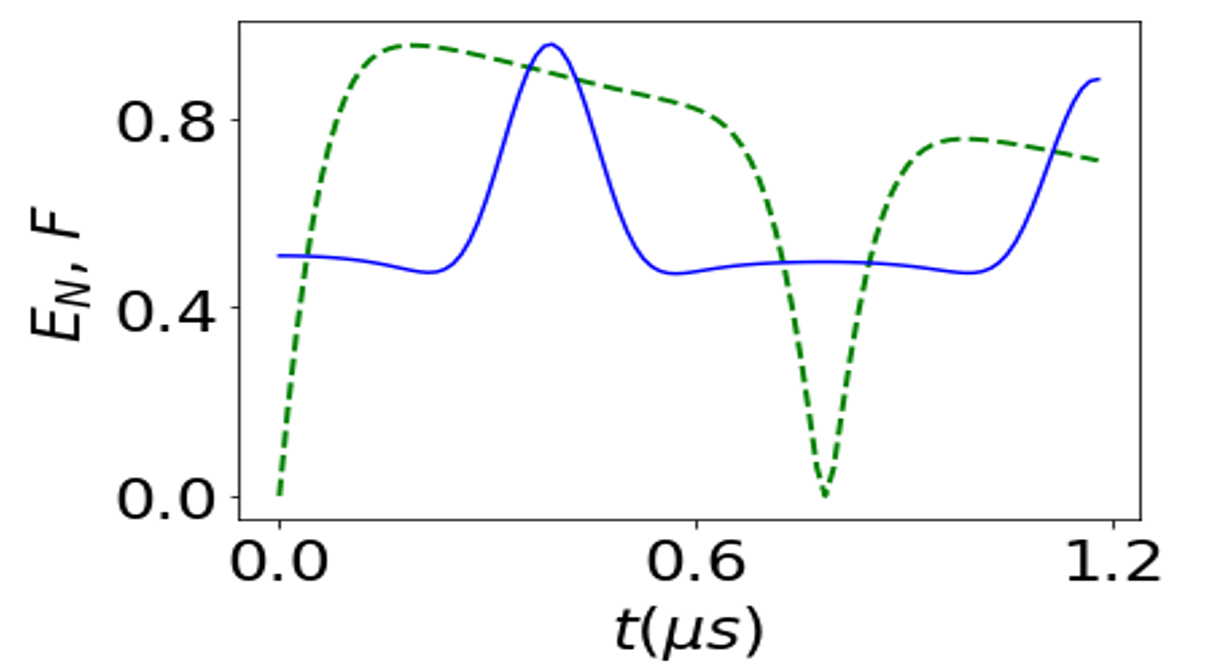}
            \caption{(Color online) Entanglement $E_N$ (green dotted line) and fidelity $F$ (solid blue line) measurement of the state $|\psi\rangle_1$ in the presence of thermal noise. As expected, the state $|\psi\rangle_1$ evolves into the Bell-cat state $|\psi\rangle_1 = (-\beta_1 \rangle|e_1\rangle + |\beta_1 \rangle|g_1\rangle)/\sqrt{2}$ at the interval $\pi/\lambda_1$, $3\pi/\lambda_1$, ... and so on. The parameters used are $\beta_1=\sqrt{2}$, $\gamma_1=0.1$MHz, $\Gamma_1=0.1$MHz, $n_{th}=0.03$, and $\lambda_1=8$MHz.}
            \label{fig:2}
        \end{figure}

    
    \section{Generation of bipartite cat state.}
        \label{sec:Bipartite cat state}
\begin{figure}[ht]
            \centering
            \includegraphics[width=0.45\textwidth]{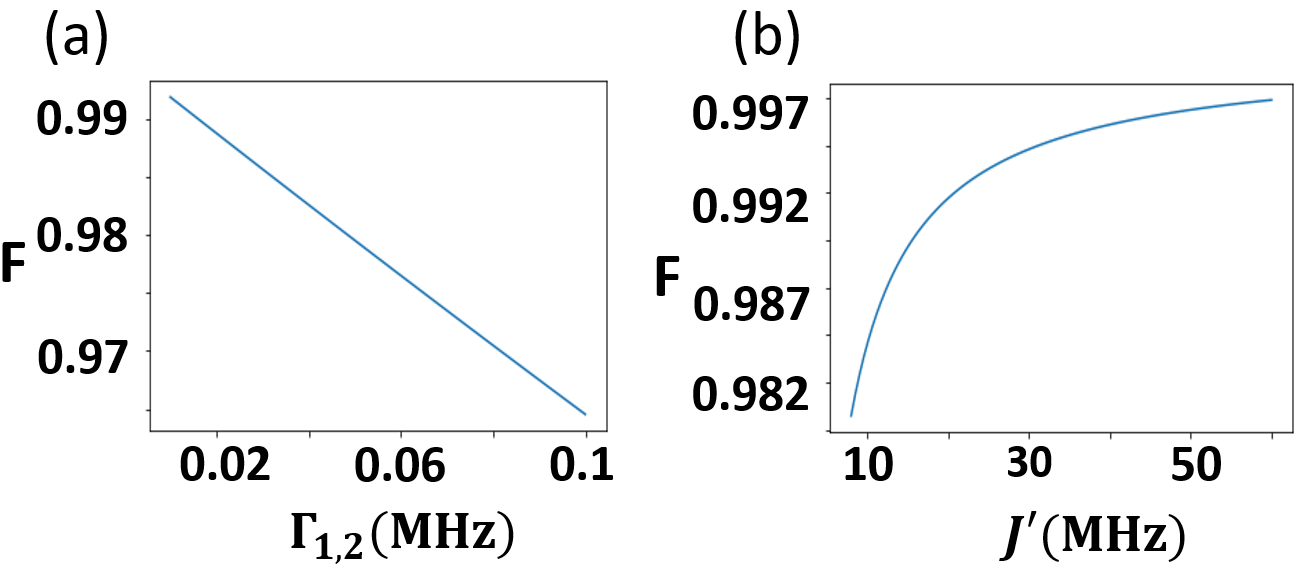}
            \includegraphics[width=0.45\textwidth]{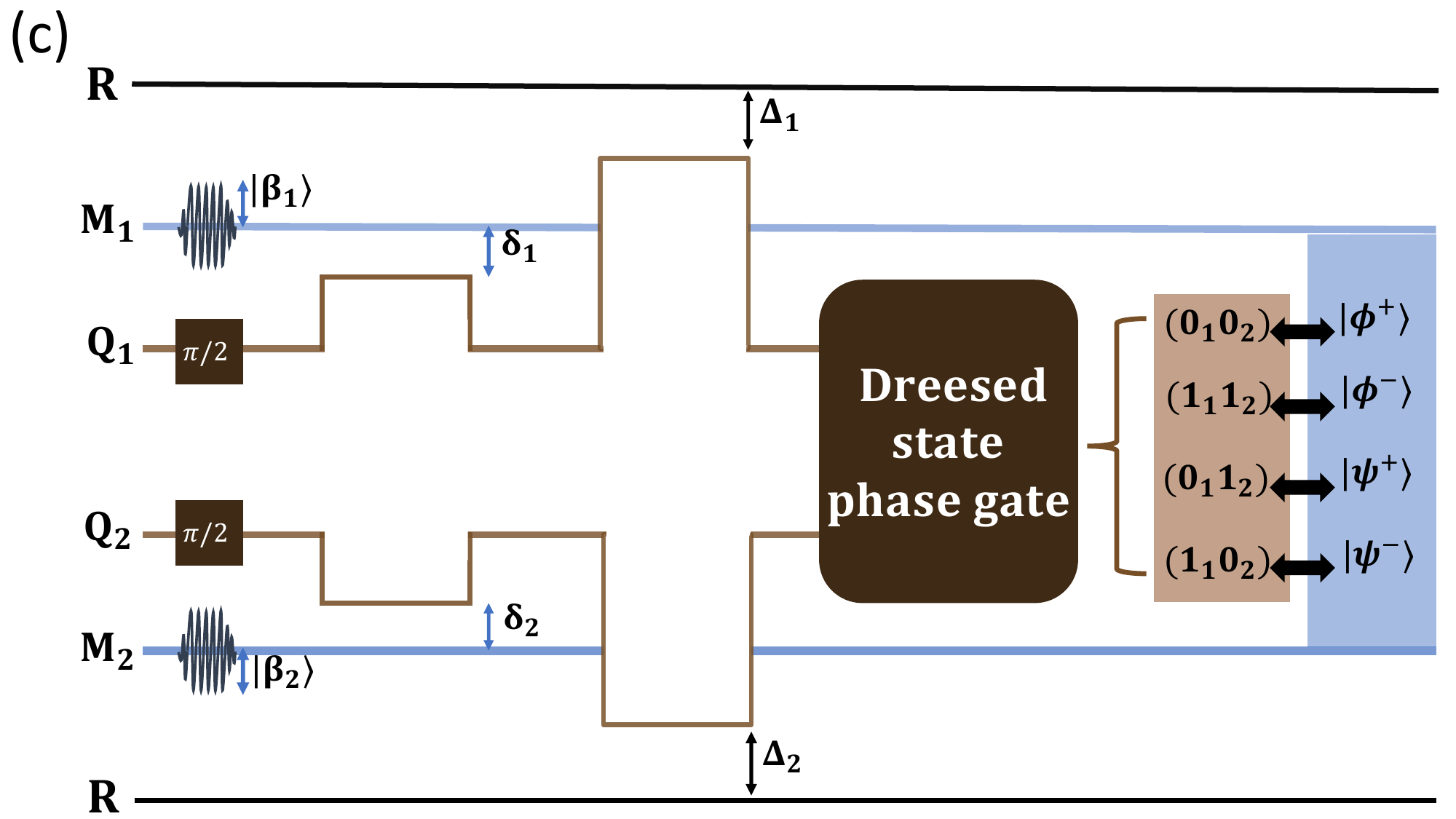}
            \caption{(Color online). The change in fidelity of the dressed-state $|++\rangle_{\tau}$ with respect to decay rate $\Gamma_{1,2}$ and coupling strength $J'$ are shown in (a) and (b), respectively. The coupling strength in (a) is $J'=8$MHz and decay rates in (b) are $\Gamma_{1,2}=0.1$MHz. (c) Sequence of qubit detuning and Bell state preparation and measurement. Both the qubits $Q_1$ and $Q_2$ are initially prepared in the superposition state by applying $\pi/2$ pulse. Similarly,  mechanical resonators ($M_1$ and $ M_2$) are initially prepared in coherent states ($|\beta_1\rangle$ and $|\beta_2\rangle$) by driving it on resonance with a microwave drive. We then initiate the qubit-mechanical dispersive interaction by detuning the qubits to $\delta_1$ and $\delta_2$ with respect to the resonators.   After time $t_1=t_2=\pi/\lambda_{1,2}$ we bring the qubits back to their idle frequencies, terminating the qubit-mechanical dispersive interaction. Now we switch on the qubit-qubit interaction by detuning the qubits to $\Delta_1$ and $\Delta_2$ with respect to the microwave resonator R. After time $\tau=\pi/2J'$, we bring the qubits back to their idle frequencies and start the Bell state measurements. During the time $\tau$, the qubits are continuously driven generating the Dressed-state phase gate.}
            \label{fig:3}
        \end{figure}

\begin{figure}[ht]
            \centering
            \includegraphics[width=0.45\textwidth]{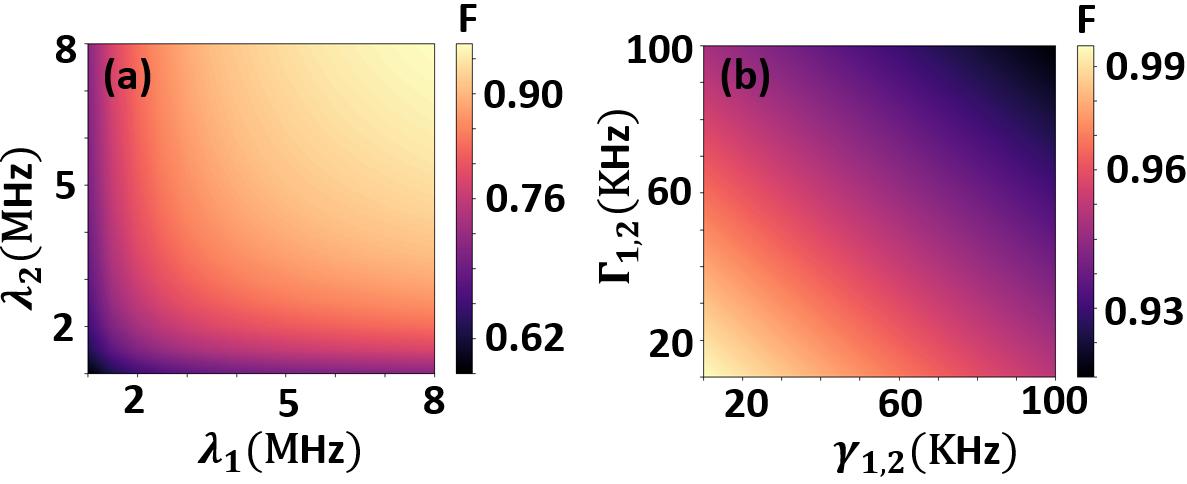}
            \includegraphics[width=0.45\textwidth]{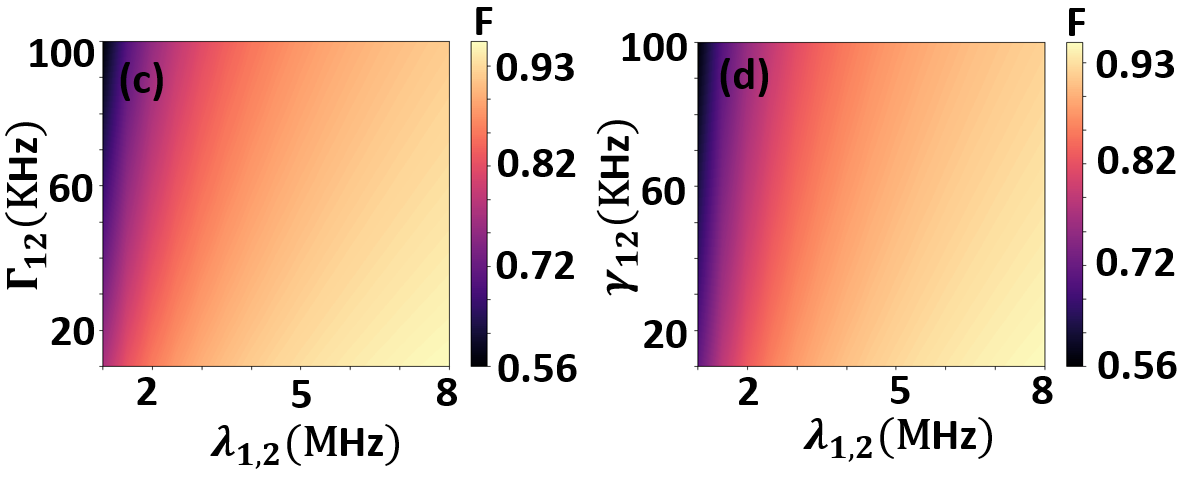}

            \caption{(Color online) Fidelity of the bipartite cat state $|C_+ \rangle$ as a function of (a) $\lambda_1$ and $\lambda_2$, (b) $\gamma_1$ and $\Gamma_1$, (c) $\Gamma_1$ and $\lambda_1$, (d) $\lambda_1$ and $\gamma_1$. The decay rates in (a) are $\gamma_{1,2}=0.1$MHz and $\Gamma_{1,2}=0.1$MHz. The coupling strength in (b) are $\lambda_{1,2}=8$MHz. In (c), the decay rates of first and second resonators are $\gamma_{1,2}=0.1$MHz. The qubits decay rate in (d) are $\Gamma_{1,2}=0.1$MHz. In all the four plots,  the variable parameters are varied simultaneously for both the qubit-mechanical systems. As expected, we observe increase in fidelity when the decay rates decreases and coupling constant increases.}
            \label{fig:5}
        \end{figure}

         After the interaction time of $t_1=\pi/\lambda_1$ $(t_2=\pi/\lambda_2)$, the state of the qubit-mechanical bipartite state (Eq.\ref{eqn:2}) become $|\psi\rangle_1 = (|-\beta_1\rangle|e_1\rangle + |\beta_1 \rangle|g_1\rangle)/\sqrt{2}$ $(|\psi\rangle_2 = (|-\beta_2\rangle|e_2\rangle + |\beta_2 \rangle|g_2\rangle)/\sqrt{2})$. The state  of the combined system $|\Psi\rangle$ is then given by the tensor product of the states of the two hybrid systems, i.e.,$|\Psi\rangle=|\psi\rangle_1|\psi\rangle_2$
        \begin{eqnarray}
            \label{eqn:4}
            |\Psi\rangle &=& \frac{1}{2}[e_1e_2-\beta_1-\beta_2\rangle+|e_1g_2-\beta_1+\beta_2\rangle \nonumber\\
            && + |g_1\,e_2\,\beta_1-\beta_2\rangle+|g_1\,g_2\,\beta_1\,\beta_2\rangle],
        \end{eqnarray}
        where $|\beta_2\rangle$ is the coherent amplitude of the second resonator. In terms of Bell's basis, $|\phi^\pm\rangle=(1/\sqrt{2})(|e_1e_2\rangle\pm i|g_1g_2\rangle$ and $|\psi^\pm\rangle=(1/\sqrt{2})(|e_1g_2\rangle\pm i|g_1e_2\rangle$, the wave function of the combined system (Eq. \ref{eqn:4}) can be rearranged to 
        \begin{eqnarray}
            \label{eqn:5}
            |\Psi\rangle &=& \frac{1}{2\sqrt{2}}[|\phi^+\rangle(|-\beta_1-\beta_2\rangle-i|\beta_1\,\beta_2\rangle) \nonumber \\
             && + |\phi^-\rangle(|-\beta_1-\beta_2\rangle+i|\beta_1\,\beta_2\rangle)  \\
            && + |\psi^+\rangle( |-\beta_1\,\beta_2\rangle -i|\beta_1-\beta_2\rangle ) \nonumber \\
            && + |\psi^-\rangle( |-\beta_1\,\beta_2\rangle +i|\beta_1-\beta_2\rangle ) \nonumber].
        \end{eqnarray}
        Therefore, by measurement the Bell's states $|\phi^\pm\rangle$ and $|\psi^\pm\rangle$, the two mechanical resonators are projected into the bipartite cat states $|C'_\mp\rangle=\mathcal{N}'_\mp(|-\beta_1-\beta_2\rangle\mp i|\beta_1\,\beta_2\rangle)$ and $|C_\mp\rangle=\mathcal{N}_\mp( |-\beta_1\,\beta_2\rangle \mp i|\beta_1-\beta_2\rangle )$, respectively. The Bell state measurement of the two qubits is performed by first switching on the interaction between them ($\hbar J'(\hat{\sigma}_1^-\hat{\sigma}_2^+ + \hat{\sigma}_1^+\hat{\sigma}_2^-)$, refer appendix \ref{app:ham_dispersive}). This interaction, which will entangle the two qubits, is achieved by red detuning the second qubit from the cavity and blue detuning the same qubit from the second resonator. Such detuning arrangement or dispersive coupling allows the exchange of virtual cavity photons with the qubits. The virtual interaction between the two qubits only produces two of the four Bell states, $|\psi^\pm\rangle$. Instead of transforming from one Bell state to the other for every measurement, we generate all four of the Bell state simultaneously. To generate all four Bell states, we continuously drive the two qubits, resulting in two dressed states $|\pm\rangle_j = (1/\sqrt(2))(|g\rangle_j\pm e^{i\Phi_j}|e\rangle_j)$, $j=1,2$. The qubit cannot go into transitions between different dressed states under the conditions $\Phi_1=\Phi_2=\Phi$ and $|A_1-A_2|>>|J'|$, where $A_j$ and $\Phi_j$ are the drive amplitude and phase. When the phases of both the driving fields are reversed right in the middle of the
         two-qubit interaction time, then the dressed state of the two qubit evolves to $|++\rangle_\tau=(1/2)(i|\phi^-\rangle + |\phi^+\rangle + i|\psi^-\rangle + |\psi^+\rangle)$, where $\tau = \pi/(2J')$. The above protocol for producing the dressed state $|++\rangle_\tau$ is known as  dressed-state phase gate. In Fig. \ref{fig:3}(a)(b), we plot the fidelity of the dressed state generated through the dressed-state phase gate. The measurement fidelity of the Bell state is dependent on the fidelity of the dressed-state phase gate.
        The four Bell state generated through the dressed-state phase gate can be mapped onto the computational basis as $|\phi^+\rangle\rightarrow i(0_10_2)$, $|\phi^-\rangle\rightarrow(1_11_2)$, $|\psi^+\rangle\rightarrow i(0_11_2)$, and $|\psi^-\rangle\rightarrow(1_10_2)$ \cite{PhysRevLett.123.060502, PhysRevLett.121.130501} (appendix \ref{Bell state measurement}). Therefore, based on the measurement outcomes of the two qubits in the computational basis, all four bipartite cat states of the resonators are generated. The sequence of operations performed in our scheme to generate the bipartite cat state is illustrated in Fig. \ref{fig:3}(c).
        \\
        The above protocol is realized in the noisy environment by first independently evolving the two qubit-mechanical hybrid systems under the Lindblad master equation and then performing the projective measurement on the density matrix of the combined system, $\hat{\rho}=\hat{\rho}_1\hat{\rho}_2$. The reduced density matrices of the four bipartite Bell-cat states after the projective Bell state measurement on the two qubits read as,
        \begin{eqnarray}
            \label{eqn:6}
                \hat{\rho}_{C'_{\mp}} &=& \frac{\langle\phi^\pm|\hat{\rho}|\phi^\pm\rangle}{tr(\langle\phi^\pm|\hat{\rho}|\phi^\pm\rangle)},\nonumber\\
                &&
                \hspace{-1cm}\hat{\rho}_{C_{\mp}} = \frac{\langle\psi^\pm|\hat{\rho}|\psi^\pm\rangle}{tr(\langle\psi^\pm|\hat{\rho}|\psi^\pm\rangle)}.
        \end{eqnarray}
        Constructing the reduced density matrix (Eq. \ref{eqn:6}) on the number basis will require a huge subspace, and performing joint state tomography on the resonators will be experimentally challenging. So, we reconstruct the density matrix into a two-level system subspace by projecting into the basis state ($|\beta_{1,2}\rangle$, $|-\beta_{1,2}\rangle$). The Pauli operators for the resonator can be obtained by measuring the displaced phonon number parity observable $\hat{P}_{\beta_1}=\hat{D}_{\beta_1}\hat{P}\hat{D}^{\dagger}_{\beta_1}$, where $\hat{D}_{\beta_1}$ is the displacement operator and $\hat{P}$ the phonon number parity operator \cite{NatCommun.6.8970(2015), Nat.Phys.18.794–799(2022)}. The Pauli operators for the resonator mode become
        \begin{eqnarray}
        \label{eqn:7}
            \hat{X}_{\beta_1}&=&\hat{P}_0, \hspace{0.5cm} \hat{I}_{\beta_1} = [\hat{P}_{\beta_1}+\hat{P}_{-\beta_1}], \nonumber\\
            &&
            \hspace{-1cm}\hat{Y}_{\beta_1} = \hat{P}_{\frac{-i\pi}{8\beta^*_1}}, \hspace{0.5cm} \hat{Z}_{\beta_1} = [\hat{P}_{\beta_1} - \hat{P}_{-\beta_1}] .
        \end{eqnarray}
        We have assumed a large orthonormal cat state, i.e., $\langle\beta_1|-\beta_1\rangle<<1$. Four Wigner functions ($W(\alpha)=\frac{2}{\pi} \langle P_{\alpha}\rangle$, where $\alpha=0,\beta,-\beta, -i\pi/8\beta^*$) are required to reconstruct the state. The basis for the Pauli operators is ($|\beta_1\rangle$, $|-\beta_1\rangle$) which is analogous to the qubit basis ($|e\rangle$, $|g\rangle$). Similar Pauli operators and basis ($|\beta_2\rangle$, $|-\beta_2\rangle$)  can be generated for the second resonator using the same approach.
        \begin{figure}[ht]
            \centering
            \includegraphics[width=0.45\textwidth]{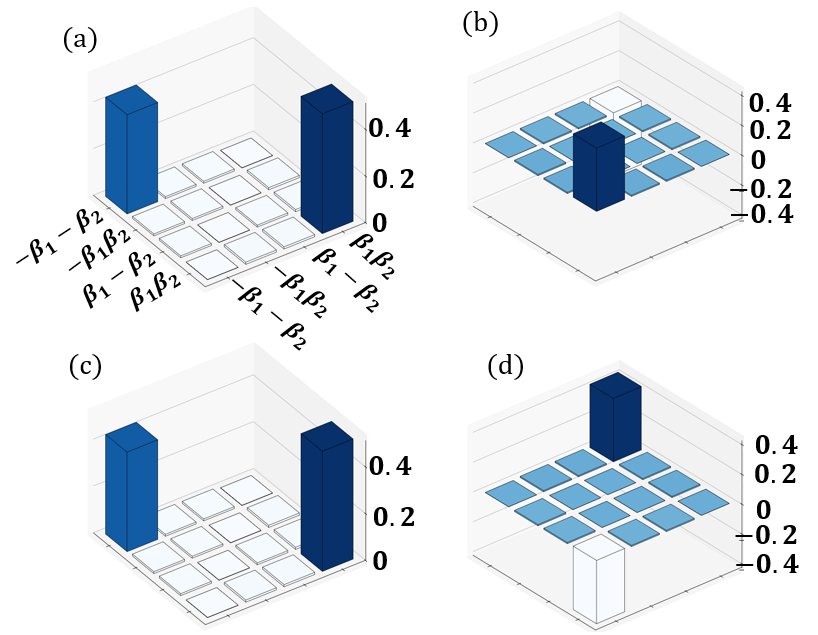}
            \includegraphics[width=0.45\textwidth]{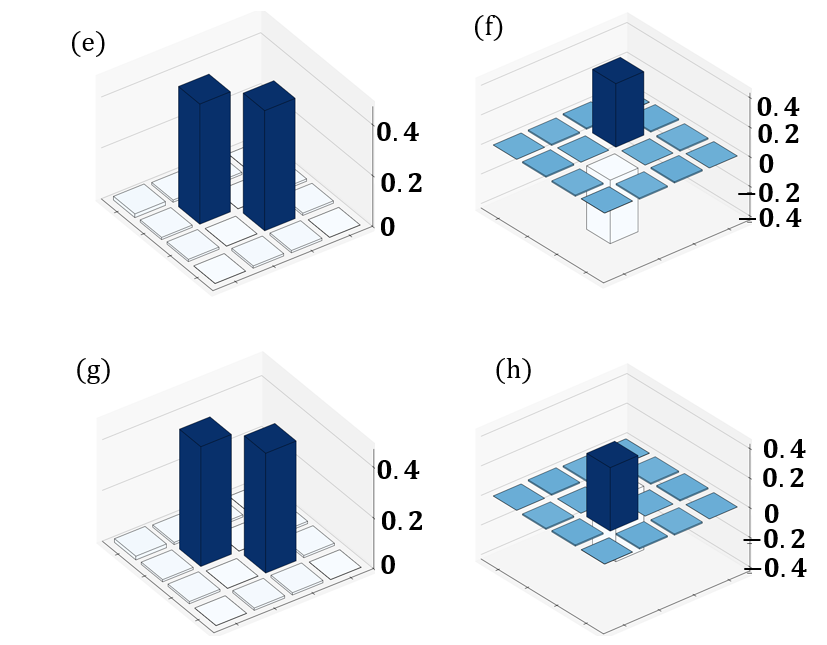}
            \caption{(Color online) Construction of the four bipartite phononic cat states density matrices in the two-level subspace. (a) and (b) ((c) and (d)) represent the real and imaginary parts of $\hat{\rho}_{C'_-}$ ($\hat{\rho}_{C'_+}$), respectively. Similarly, (e) and (f) ((g) and (h)) represent the real and imaginary parts of $\hat{\rho}_{C_-}$ ($\hat{\rho}_{C_+}$), respectively. The density matrices resemble those of the two-qubit Bell state \cite{PhysRevLett.123.060502}. We have used the resonator coherent amplitudes $\beta_{1,2}=\sqrt{2}$. }
            \label{fig:4}
        \end{figure}
        In Fig. \ref{fig:4}, we plot the joint density matrix of the two resonators in the joint basis states ($|\beta_1\,\beta_2\rangle$, $|-\beta_1\,\beta_2\rangle$, $|\beta_1\,-\beta_2\rangle$, and $|-\beta_1\,-\beta_2\rangle$). We observe four bipartite cat states having fidelities $F_{C'_-}=0.919$, $F_{C'_+}=0.919$, $F_{C_-}=0.92$, and $F_{C_+}=0.92$ and entanglements $E_{C'_-}=0.799$, $E_{C'_+}=0.799$, $E_{C_-}=0.799$, and $E_{C_+}=0.799$. The dip in the density matrix element is mainly attributed to the relaxation and decoherence effect of the two qubits as well as the relaxation of the two resonators. As shown in Fig. \ref{fig:5}, the fidelity of the bipartite cat state can be significantly improved by improving the decay rates of both the qubit and the phononic resonator. For example, we get $F_{C'_-}=0.9581$, $F_{C'_+}=0.9581$, $F_{C_-}=0.9581$, and $F_{C_+}=0.9581$ and entanglements $E_{C'_-}=0.895$, $E_{C'_+}=0.895$, $E_{C_-}=0.895$, and $E_{C_+}=0.895$ for decay rates $\gamma_{1,2}=0.05Mhz$ and $\Gamma_{1,2}=0.05MHz$. In Fig. \ref{fig:5}, we have generated the fidelity variation for $|C_+\rangle$ state. We get similar plots for the other cat states too. In addition to the decays resulting from direct contact with the noisy environment, the other factors that lead to the infidelities of the prepared state include read-out errors while measuring the mechanical resonators and errors while performing the Bell measurement. 
        The four bipartite cat states resemble the traditional qubit Bell states generated using a Hadamard and CNOT gate in a quantum circuit. Going by this similarity, the scheme proposed here could be used to implement an entanglement gate for a continuous variable resonator qubit, and may find applications in the bosonic based quantum processors.


    \section{Bell's test of the resonator bipartite cat state}
        \label{sec:sync}
\begin{figure}[ht]
            \centering
            \includegraphics[width=0.45\textwidth]{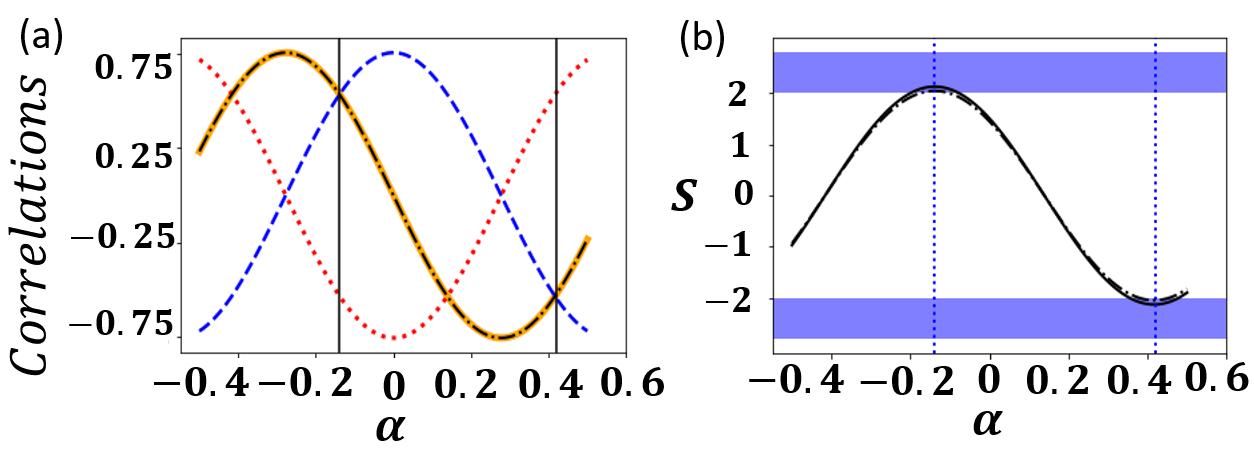}
            \includegraphics[width=0.23\textwidth]{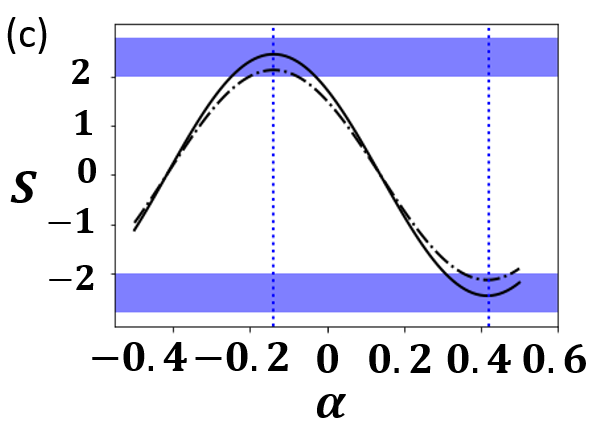}

            \caption{(Color online) Measurement correlations and inequality test are conducted with respect to the change in the displacement amplitude $\alpha$ for mechanical coherence strength $\beta_{1,2}=\sqrt{2}$. In (a), the expectation values of all the joint measurement correlations between the two resonators are depicted. The red dotted, blue dashed, black dashed-dotted, and solid orange lines represent the measurement correlations $\langle Y_\alpha X_2\rangle$, $\langle X_\alpha Y_2\rangle$, $\langle Y_\alpha Y_2\rangle$, and $\langle X_\alpha X_2\rangle$, respectively.
            In (b), the CHSH value $S$ of the Bell inequality is shown. The solid and dashed-dotted lines correspond to the coupling strengths $\lambda_1 = \lambda_2 = 8$ MHz and $\lambda_1 = \lambda_2 = 7$ MHz, respectively. For both cases, the maximum $|S|$ values occur at $\alpha_1 = -0.14$ and $\alpha_2 = 0.41$, denoted by the two dotted vertical lines, with corresponding values of $2.132$ and $2.049$. The decay rates in (b) are $\gamma_{1,2}=0.1Mhz$ and $\Gamma_{1,2}=0.1MHz$. For different decay rates $\gamma_{1,2}=0.05Mhz$ and $\Gamma_{1,2}=0.05MHz$, we get $|S|=2.454$ for $\lambda_1 = \lambda_2 = 8$ MHz, as shown in (c) (solid line). The dashed-dotted line in (c) is for  $\gamma_{1,2}=0.1Mhz$, $\Gamma_{1,2}=0.1MHz$ and $\lambda_1 = \lambda_2 = 8$ MHz, which is similar to the solid line in (b). The shaded horizontal lines represent the regions of the quantum limit.}
            \label{fig:6}
        \end{figure}

        The bipartite entangled cat states of the two resonators generated on a conditional Bell state measurement of the two qubits can be used as a platform to test the Bell's inequality in a macroscopic quantum system. Here, we perform this test by calculating the expectation values of all the correlations of the measurement outcomes measured locally at the two resonators and then determine the Clauser–Horne–Shimony–Holt (CHSH) value $S=\langle X_1X_2\rangle + \langle X_1Y_2\rangle - \langle Y_1X_2\rangle+\langle Y_1Y_2\rangle$, where, $X_1$ ($X_2$) and $Y_1$ ($Y_2$) are the observables of the first (second) resonator. The observables are measured in the resonator qubit subspace discussed above. As per CHSH inequality, a system is said to be classically correlated if $|S|\le2$ and quantumly if $2<|S|\le2\sqrt{2}$. The correlation of the observables is measured by first choosing two arbitrary values of $\beta'_1$ and $\beta'_2$ corresponding to  $X_1$, $Y_1$ and $X_2$, $Y_2$, respectively. We then rotate the resonator detector basis by coherently displacing the observables $X_1$ and $Y_1$ to $X_\alpha=D_{-i\alpha}X_1D_{i\alpha}$ and 
        $Y_\alpha=D_{-i\alpha}Y_1D_{i\alpha}$ or
        \begin{eqnarray}
        \label{eqn:8}
            X_\alpha &=& X_1\, cos2(\alpha\beta'^*_1+\alpha^*\beta'_1) +Y_1sin2(\alpha\beta'^*_1+\alpha^*\beta'_1),\nonumber\\
            &&
            \hspace{-1cm}Y_\alpha = Y_1\, cos2(\alpha\beta'^*_1+\alpha^*\beta'_1)-X_1sin2(\alpha\beta'^*_1+\alpha^*\beta'_1).
        \end{eqnarray}
        Here, $\alpha$ is the coherent displacement amplitude of the resonator. By changing the amplitude $\alpha$, we are able to rotate the measurement basis direction and perform measurements at all possible orientations of the detectors. The $|S|$ value for the state $|C_+\rangle$ is shown in Fig. \ref{fig:6}(b). We observe maximum $|S|$ value when $\beta'_{1,2}=\beta_{1,2}$ and at the displacement amplitude $\alpha_1=-0.14$ and $\alpha_2=0.41$. Furthermore, the $\alpha_1$, $\alpha_2$ and $\beta'_1$ amplitudes corresponding to the maximum $|S|$ value satisfy the relations $4\alpha_2\beta'_1=3\pi/4$ and $4\alpha_1\beta'_1=-\pi/4$. 
        The corresponding observables (Eq. \ref{eqn:8}) at these amplitudes become $X_{\alpha_1}=(X_1-Y_1)/\sqrt{2}$ and $Y_{\alpha_1}=(X_1+Y_1)/\sqrt{2}$. Therefore, the $|S|$ value observed in the figure exceeds the classical bound limit and attains a value which is less than the ideal quantum bound limit $2\sqrt{2}$ (see Fig \ref{fig:6}(b)(c) . As we decrease the decay rates of the resonators and the qubits, the maximum attainable $|S|$ value also increases, as shown in Fig \ref{fig:6}(c). The expectation values of all the measurement correlations are also shown in Fig. \ref{fig:6}(a). The behaviour of these correlations as we change the displacement amplitude $\alpha$ resembles the one observed in two-qubit Bell test experiments \cite{Nature.617.265–270.(2023)}. By integrating two phononic crystal resonators into the experimental arrangement typically employed for conducting a loophole-free Bell test with two superconducting qubits \cite{Nature.617.265–270.(2023)}, our proposed approach could be employed to examine the Bell inequality of a phononic cat state. Additionally, the bipartite phononic cat state generated through entanglement swapping in this work could hold significant practical implications for the advancement of complex quantum network processors based on continuous variable resonators. 

    
    \section{Conclusion}
        \label{sec:conc}
        In conclusion, we propose a scheme to generate four bipartite phononic cat states by performing a projective Bell state measurement on two superconducting qubits. Initially uncoupled phononic crystal resonators, each coupled to a different superconducting qubit, become entangled through entanglement swapping from qubit-mechanical to mechanical-mechanical interactions. Displaced phonon parity measurement is done for generating joint density matrix of the two resonators in two-level subspace. These joint density matrices resemble those of traditional qubit Bell states generated using a Hadamard and CNOT gate in a quantum circuit. Subsequently, we investigate the Bell inequality test using the CHSH formulation. The expectation values of the measurement correlations and the $S$ values of the CHSH inequality test obtained here are akin to those observed in \cite{Nature.617.265–270.(2023)} for two superconducting qubits. The bipartite phononic
        cat state generated through entanglement swapping in
        this work may find useful in implementing quantum network processors
        based on continuous variable resonators. Furthermore, the scheme presented in this work also serves
        as a platform for studying Bell inequality test in continuous variable system.


    \section*{Acknowledgement}
        R.N. gratefully acknowledges support of a research fellowship from CSIR, Govt. of India. U.D. gratefully acknowledges a research fellowship from MoE, Government of India. A.K.S. acknowledges the STARS scheme, MoE, government of India (Proposal ID
        2023-0161).


    \appendix

    \section{Dispersive Hamiltonian}
        \label{app:ham_dispersive}
        The interaction part of the complete Hamiltonian, including the cavity resonator connecting the two qubits, is given by the Jaynes-Cummings interaction
        \begin{eqnarray}
        \label{eqn:a1}
            \hat{H}_{int} &=& g_1(\hat{b}^{\dagger}_1\hat{\sigma}_-^1+\hat{b}_1\hat{\sigma}_+^1)+g_2(\hat{b}^{\dagger}_2\hat{\sigma}_-^2+\hat{b}_2\hat{\sigma}_+^2)\nonumber\\
            &&
            \hspace{-0.8cm}+G_1(\hat{a}^{\dagger
            }\hat{\sigma}_-^1+\hat{a}\hat{\sigma}_+^1)+ G_2(\hat{a}^{\dagger}\hat{\sigma}_-^2+\hat{a}\hat{\sigma}_+^2).
        \end{eqnarray}
        Here, $g_1$ $(g_2)$ and $G_1$ $(G_2)$ are the resonant coupling strength of mechanical-qubit and qubit-cavity interactions, respectively. $\hat{a}$ and $\hat{a}^\dagger$ are the creation and annihilation operators of the cavity resonator. We blue detune the first resonator from the first qubit ($\delta_1=\omega_1-\Omega_1$), second qubit from the cavity resonator ($\Delta_2=\omega-\Omega_2$), and red detune the cavity from the first qubit ($\Delta_1=\Omega_1-\omega$), second resonator from the second qubit ($\delta_2=\Omega_2-\omega_2$). Here, $\omega$ refers to the cavity resonance frequency. Because of the detuning, we can transform the Jaynes-Cummings interaction (Eq. A1) into a dispersive one by performing a unitary operation $U=exp\{ g_1/\delta_1(\hat{b}^{\dagger}_1\hat{\sigma}_-^1-\hat{b}_1\hat{\sigma}_+^1) + g_2/\delta_2(\hat{b}_2\hat{\sigma}_+^2-\hat{b}^{\dagger}_2\hat{\sigma}_-^2) + G_1/\Delta_1(\hat{a}\hat{\sigma}_+^1-\hat{a}^{\dagger}\hat{\sigma}_-^1) + G_2/\Delta_2(\hat{a}^{\dagger}\hat{\sigma}_-^2-\hat{a}\hat{\sigma}_+^2)\}$ on $\hat{H}_{int}$.
        \begin{eqnarray}
        \label{eqn:a2}
            \hat{H}_{dis} &=& \hbar(\eta_1\hat{\sigma}_z^1+\eta_2\hat{\sigma}_z^2)\hat{a}^\dagger\hat{a} + \hbar\lambda_1\hat{\sigma}_z^1\hat{b}_1^\dagger\hat{b}_1 + \hbar\lambda_2\hat{\sigma}_z^2\hat{b}_2^\dagger\hat{b}_2 \nonumber\\
            && 
            + \hbar J(\hat{\sigma}_1^-\hat{\sigma}_2^+ + \hat{\sigma}_1^+\hat{\sigma}_2^-) + \hbar\chi_1(\hat{a}\hat{b}_1^\dagger+\hat{b}_1\hat{a}^\dagger)\hat{\sigma}_1^z \nonumber\\
            &&
            +\hbar\chi_2(\hat{a}\hat{b}_2^\dagger+\hat{b}_2\hat{a}^\dagger)\hat{\sigma}_2^z,
        \end{eqnarray}
        where $\eta_1=G_1^2/\Delta_1$, $\eta_2=G_2^2/\Delta_2$, $\lambda_1=g_1^2/\delta_1$, $\lambda_2=g_2^2/\delta_2$,
        $J=G_1G_2(1/\Delta_1-1/\Delta_2)$, $\chi_1=g_1G_1(1/\delta_1-1/\Delta_1)$ and $\chi_2=g_2G_2(1/\Delta_2-1/\delta_2)$. The first term in the Hamiltonian A2 can be neglected since the cavity remains in the ground state. We can ignore the last three terms if we only dispersively detune the qubit-mechanical pair and the detuned qubit frequency is way off from the cavity frequency such that there is no interaction between the qubit and the cavity \cite{PhysRevLett.123.060502}. If the qubit-mechanical and qubit-cavity detuning are simultaneous, then by choosing $\Delta_1=\Delta_2$, $\delta_1=\Delta_1$ and $\delta_2=\Delta_2$, we can also neglect the last terms \cite{Nature.604.463–467(2022),Nature.449.443–447(2007)}. The remaining two interacting terms are the ones used in Hamiltonian \ref{eqn:ham_qm} and \ref{eqn:ham_om}. Thus, we see that by changing the detuning, we can independently evolve the two qubit-resonator pairs. In our scheme we first evolve the qubit-mechanical pair dispersively up to a certain time period. We then bring back the qubits to their idle frequencies so that the interaction between the qubit and mechanical resonator is turned off \cite{Nat.Phys.18.794–799(2022)}. We now initiate the qubit-qubit interaction (fourth term) by detuning back the qubit with respect to the cavity in order to perform Bell state measurement. This can be done in two ways. First, by choosing $\Delta_1 \neq \Delta_2$, or second, by red detuning the second qubit from the cavity and blue detuning the second qubit from the second resonator. In the second approach the coupling constant $J$ becomes $J'=G_1G_2(1/\Delta_1+1/\Delta'_2)$, where $\Delta'_2=\Omega_2-\omega$. The coupling sequence is shown in the Fig. \ref{fig:3}(d).

        \section{Bell State of the two qubits}
        \label{Bell state measurement}
        In order to distinguish all the Bell states of the two qubit, we resonantly drive the qubits individually. 
        In the interaction frame, the Hamiltonian of the two qubit inteaction is 
        \begin{equation}
        \label{Bell int}
            \hat{H} = \hbar (J'\hat{\sigma}_1^+\hat{\sigma}_2^- +
             \sum_{j=1,2}A_je^{-i\Phi_j}\sigma_j^+)+H.c.
        \end{equation}
        Here, $A_j$ and $\Phi_j$ are the Rabi frequency and phase, respectively, of the drive applied to the qubits. We have assumed that $J'>>\lambda_1,\lambda_2$. The drive produces two dressed states $|\pm\rangle_j = (1/\sqrt(2))(|g_j\rangle\pm e^{i\Phi_j}|e\rangle_j)$. The qubit cannot go into transitions between different dressed states under the conditions $\Phi_1=\Phi_2=\Phi$ and $|A_1-A_2|>>|J'|$. Then the Hamiltonian \ref{Bell int} in the dressed state basis reduces to \cite{PhysRevLett.121.130501, PhysRevLett.123.060502}.
        \begin{equation}
            \hat{H}_{eff} = \frac{1}{2}\hbar J'S_{z1}S_{z2} + \hbar \sum_{j=1,2}A_jS_{zj}.
        \end{equation}
         Here, $S_zj=|+\rangle_j\langle+|_j - |-\rangle_j\langle-|_j$. If the phases of both the driving fields are reversed right in the middle of the
         two-qubit interaction time, then the dressed state $|+\rangle_1 |+\rangle_2$ evolves to $|++\rangle_t = exp(iJ't/2)|+\rangle_1 |+\rangle_2$. At time $\tau=\pi/2J'$, the dressed state become $|++\rangle_\tau=(1/2)(i|\phi^-\rangle + |\phi^+\rangle + i|\psi^-\rangle + |\psi^+\rangle)$. In the computational basis ($|0\rangle$, $|1\rangle$), the state $|++\rangle_\tau$ can be obtained by applying the unitary operator
         \begin{equation}
             U=\frac{1}{\sqrt{2}}\begin{bmatrix}
            1 & 0 & 0 & i\\
            0 & 1 & i & 0\\
            0 & i & 1 & 0\\
            i & 0 & 0 & 1
            \end{bmatrix}
         \end{equation}
         Therefore, in the computational basis the Bell state $|\phi^+\rangle$ is mapped to $|0_10_2\rangle$, i.e., $i|0_10_2\rangle= U|\phi^+\rangle$. Similarly, $|\phi^-\rangle$, $|\psi^+\rangle$, and $|\psi^-\rangle$ are mapped onto $|1_11_2\rangle$, $i|0_11_2\rangle$, $|1_10_2\rangle$.

\bibliography{reference}

\end{document}